# A RECONFIGURABLE IMPEDANCE MATCHING NETWORK EMPLOYING RF-MEMS SWITCHES

*M. Bedani[*], F. Carozza[*], R. Gaddi[*], A. Gnudi[*], B. Margesin[#], F. Giacomozzi[#]*

[*] ARCES, University of Bologna, Viale Risorgimento 2, 40136 Bologna, Italy.
tel: +390512093049 fax: +390512093779. mbedani@arces.unibo.it
[#] Fondazione Bruno Kessler – irst, Via Sommarive 18, 38050 Povo (TN), Italy.

## ABSTRACT

We propose the design of a reconfigurable impedance matching network for the lower RF frequency band, based on a developed RF-MEMS technology. The circuit is composed of RF-MEMS ohmic relays, metal-insulator-metal (MIM) capacitors and suspended spiral inductors, all integrated on a high resistivity Silicon substrate. The presented circuit is well-suited for all applications requiring adaptive impedance matching between two in principle unknown cascaded RF-circuits. The fabrication and testing of a monolithic integrated prototype in RF-MEMS technology from ITC-irst is currently underway.

## 1. INTRODUCTION

The problem of impedance matching is probably the most widely investigated and yet always challenging one among the issues of radiofrequency engineering. While its fundamental theory stands on solid pillars, the day-by-day design always poses further requirements pushing for higher integration, lower losses and reusability. Microelectromechanical Systems (MEMS) technology, gradually characterizing itself as an important technology enabler for future generation radiofrequency circuits, has unsurprisingly appeared in recently proposed reconfigurable matching network solutions for RF to microwave frequencies [1], [2]. While the design of distributed microwave circuits has found a natural extension in low-loss substrate RF-MEMS technology solutions, integrated lumped-element RF-MEMS circuits for the lower RF band are still relatively rare, possibly due to the general need for relatively large and low-loss passive components integrated with reliable ohmic-contact based relays.

The presented design implements a general purpose impedance matching network, aiming at realizing a wide coverage of possible complex impedances that could be presented to an unknown RF circuit, for the lower RF frequency band, integrated on a high resistivity Silicon substrate. The design is based on a developed RF-MEMS technology of ITC-irst [3], which integrates ohmic contact-based relays, metal-insulator-metal (MIM) capacitors and suspended spiral inductors, providing a flexible design platform which is exploited in full reaching a relatively high level of circuit complexity giving the large number of degrees of freedom necessary for full-span reconfigurability.

## 2. CIRCUIT DESCRIPTION

The MEMS devices used in the presented circuit are based on ohmic switches and MIM capacitors, connected in a topology to implement varactors switching between two alternative values; both shunt and series configuration of the varactor are available. The technology process utilized for the fabrication is based on surface micromachining with an electroplated suspended gold membrane layer, one polysilicon layer for the actuation electrodes and a TiN-Ti-Al-TiN-Ti multilayer for the RF-signal path [3].

The circuit topology of the designed reconfigurable matching network is shown in Fig. 1, where two stages can be identified (for component values refer to Table 1). The first stage consists of four CL-sections (i.e. a shunt capacitor and a series inductor), each one containing two MEMS devices, therefore providing for a total amount of $2^8$ impedance configurations (8 switches). Each shunt capacitor is a two-valued MEMS varactor, while each series inductor can be partially resonated-out by a series MEMS varactor.

| Capacitances | $C_{pa}$ | $C_p$ | $C_s$ |
|---|---|---|---|
| Values (pf) | 4.5 / 6.5 | 4 / 7 | 4 / 7 |
| Capacitances | C1 | C2 | $C2_{var}$ |
| Values (pf) | 5.14 | 2.12 | 2.12 / 5.5 |
| Capacitances | $C_{decoup}$ | $C_{phase1}$ | $C_{phase2}$ |
| Values (pf) | 6 | 0.57 / 3.14 | 2 / 7.14 |
| Inductances | $L_s$ | $L_h$ | $L_{res}$ |
| Values (nH) | 16 | 8.5 | 16 |

**Table 1 : Table of component values.**

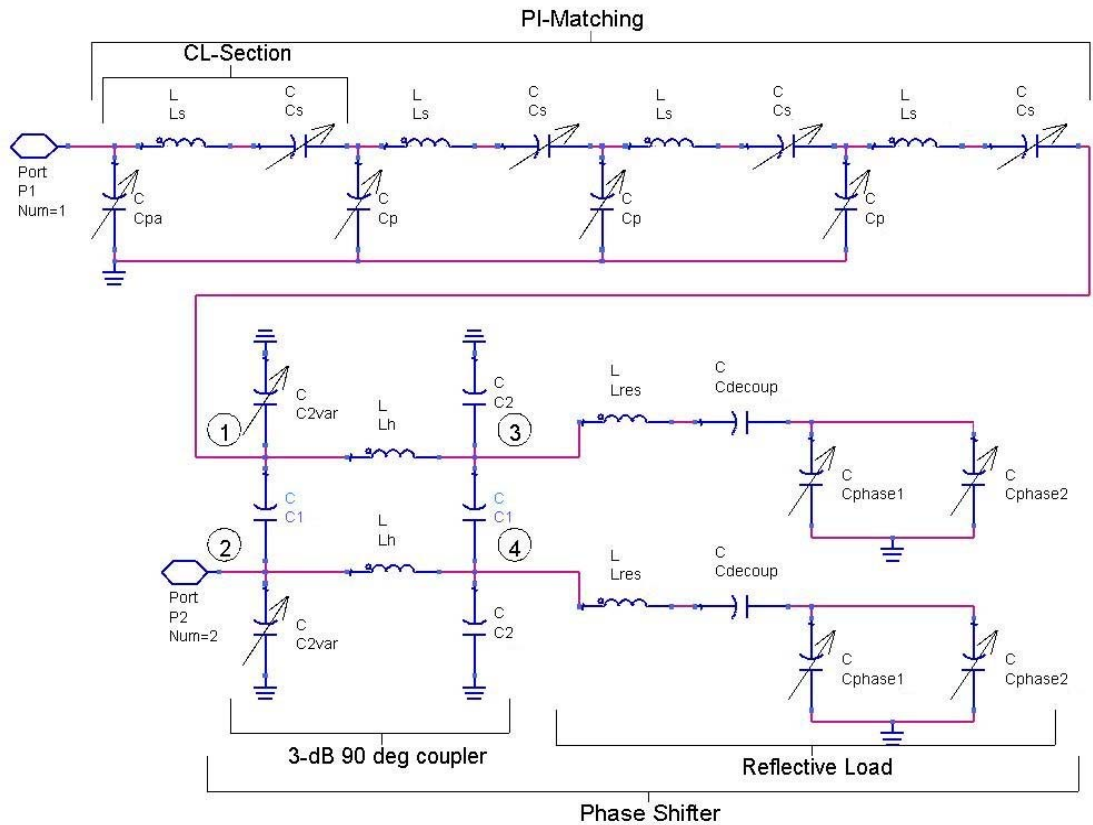

**Figure 1 : Circuit topology of the reconfigurable matching network.**

This first stage, realizing a classic Π-matching, is designed to homogeneously cover a quite large portion of the Smith chart, at the frequency of 620 MHz. The second stage is a 3-dB 90-degree coupler connected to a reflective load [4]. From the ideal design equations, when port 3 and port 4 of the 3-dB coupler (as shown in Fig. 1) are connected to an appropriate reflective load, the network allows full transmission between port 1 and port 2 while introducing a phase shift. Phase control is enabled by varying the impedance of the reflective load, in our case using MEMS varactors as shunt capacitors. Since impedances at port 3 and port 4 must be equal, the reflective load networks are identical (two shunt varactors in each one), and share the same MEMS actuation DC path. The design is tailored for a center value of the varactor capacitance corresponding to an impedance of 50 Ω. Meanwhile, $C_{min}$ and $C_{max}$ are chosen to achieve a capacitance control ratio ($C_{max}/C_{min}$) of 4. The phase control range is significantly increased by resonating the capacitance of the varactor, with the series inductor $L_{res}$. Moreover, introducing two more shunt varactors, sharing the actuation, also in the 3-dB coupler (as it can be seen from Fig. 1) adds one further degree of freedom.

The overall phase control range obtained by the second stage is about 340 degrees, applying $2^3$ different phase rotations to the impedance values synthetized by the first stage.

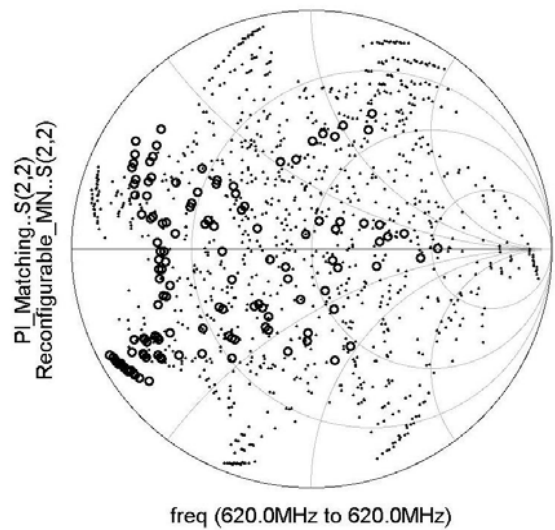

**Figure 2 : Simulated impedance values (circles: first stage; dots: full reconfigurable matching network).**

The complete network provides for a total amount of 2048 values of impedance, achieving a good coverage of the Smith chart, compared to the first stage only. ADS simulations of Smith chart coverage in terms of S22 of both the first stage (Π-matching) and the full reconfigurable matching network are shown in Fig. 2.

## 3. LAYOUT SYNTHESIS

The layout of the proposed circuit, requiring a chip area of about 40 mm$^2$ is shown in Fig. 3. Both RF port 1 and port 2 are made available for direct on-wafer testing as two GSG ports, while the actuation pads for all MEMS devices are suitable both for direct on-wafer actuation, or for wire-bonding connection following a chip-on-board assembly approach, as required due to the large number of control voltages to be independently provided. Observing the layout of Fig. 3, the cascaded networks implementing the 3-dB coupler, the reflective load and the four CL-sections (with shunt and series MEMS varactors), forming the Π-matching stage, can be distinguished.

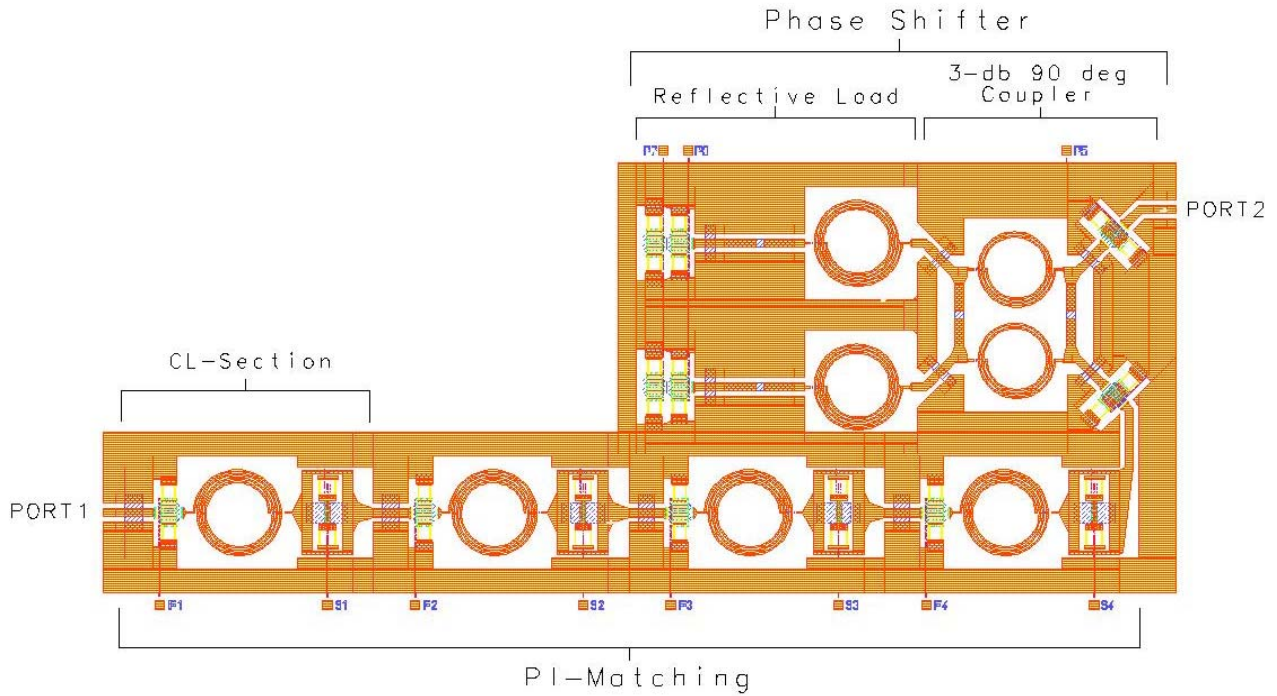

**Figure 3 : Layout of the circuit as implemented in ADS**

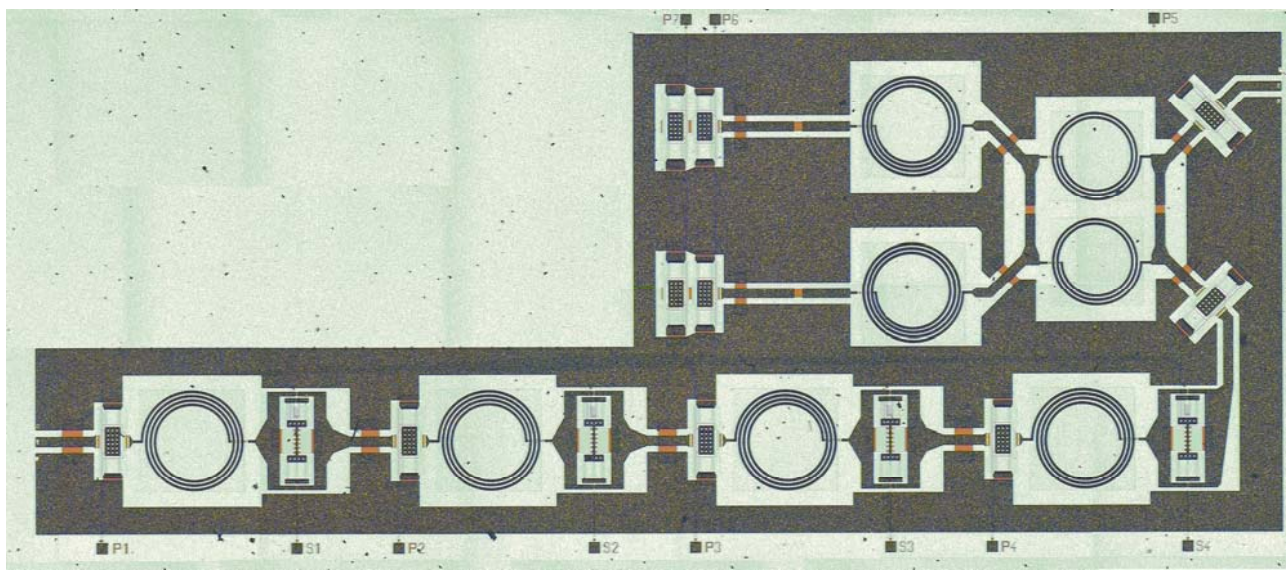

**Figure 4 : Photograph of the fabricated circuit**

Fabrication of a prototype of the designed circuit has been accomplished, while RF testing is currently under way. A picture of the complete monolithic RF-MEMS circuit in irst technology is shown in Fig. 4.

## 4. DISCUSSION

The effect of real losses related to both fixed passive elements (suspended spiral inductors and MIM capacitors) and RF-MEMS ohmic relays (shunt and series) is expected to reduce the radius of the overall reflection coefficient coverage of Fig. 2 by a factor of 0.9 [5]. Furthermore, a real assessment of the circuit overall losses will require the characterization of both input and output matching (S11 and S22) and transmission coefficient (S21). RF characterisation of each circuit element, also available as single device test structure, will allow for the real estimation of the dominant loss mechanisms, in order to optimize the overall efficiency of the proposed design approach.

To date, no measured data about the Smith chart coverage achieved by the realized circuit is available. Nonetheless, from a functional point of view, electro-mechanical characterization of the single RF-MEMS core device included in the circuit was performed, through contact-less profilometry technique based on optical interferometry. Figure 5 reports a static profile of the intrinsic ohmic switch device that realizes the two-valued varactor $C_{pa}$ (refer to the schematic shown in Figure 1). One can clearly identify the RF signal path, the suspended membrane (in 5.0 μm thick gold), and the 165 μm long anchor beams connecting the membrane to the two MIM capacitors at each side, both in shunt to ground configuration. When no actuation voltage is applied, the device is in up-state and only a small coupling parasitic capacitance affects the RF signal. On the other hand, when the device is actuated, the membrane collapses down realizing the ohmic contact between the RF signal path and the MIM capacitor to ground.

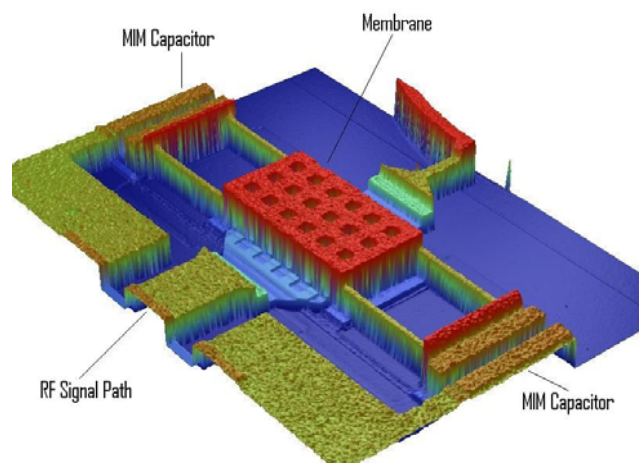

**Figure 5 : 3D static profile of a MEMS shunt varactor measured using optical interferometry.**

## 5. CONCLUSIONS

The issue of reconfigurable impedance matching has been tackled by adopting an RF-MEMS based approach: RF-MEMS ohmic switches are utilized to implement two-valued varactors, both in shunt and series configuration, which are included in a complex matching network designed to achieve a full-span reconfigurability at the frequency of 620 MHz. The circuit, providing for a total amount of 2048 impedance configurations, is composed by a classic Π-matching stage, and a 3-dB coupler which applies different phase shifts to the coverage obtained by the Π-matching stage.

The implementation of a Silicon monolithic circuit was accomplished through an RF-MEMS technology from irst.

RF testing of the fabricated prototype is currently underway. Losses related to passive components and to RF-MEMS relays are expected to be a critical aspect in the efficiency of the proposed approach.

## 6. ACKNOWLEDGEMENTS

This work was partially funded by the Italian Ministry of Research through the PRIN project "Modeling, design and characterization of MEMS devices for reconfigurable radio-frequency transceiver architectures" (2005091729).